\documentclass[a4paper,fleqn,usenatbib]{mnras}

\usepackage{latexsym}
\usepackage{amsmath}
\usepackage{amssymb}
\usepackage{fnpos}
\usepackage{hyperref}
\usepackage{tabularx}
\usepackage{xspace}
\usepackage{enumerate}
\usepackage{scalerel}
\usepackage{graphicx}
\usepackage{url}
\usepackage{caption}
\usepackage{gensymb}
\usepackage{longtable}
\usepackage{lscape}
\usepackage[normalem]{ulem} 
\usepackage[dvipsnames]{xcolor} 
\usepackage{wasysym} 
\definecolor{darkgreen}{rgb}{0.0,0.55,0.0}
\definecolor{darkblue}{rgb}{0.0,0.0,0.5}

\newcommand{\nodata}{\ensuremath{\cdots}}

\newcommand{\eal}[2]{\ifmmode{\mathrm{#1\,#2}}\else{#1\textsc{$\,$\lowercase{#2}}}\fi\xspace}
\newcommand{\feal}[2]{\ifmmode{\mathrm{#1\,#2}}\else{[#1\textsc{$\,$\lowercase{#2}}]}\fi\xspace}
\newcommand{\hfeal}[2]{\ifmmode{\mathrm{#1\,#2}}\else{#1\textsc{$\,$\lowercase{#2}}]}\fi\xspace}

\newcommand{\orcid}[1]{$^{\rm \href{https://orcid.org/#1}{\includegraphics[height=0.6em]{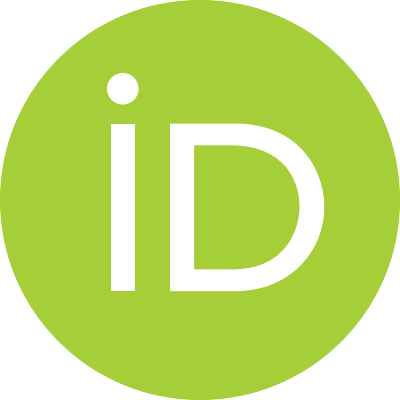}}}$}

\title[The upcoming eruption of Nova T CrB]{The First Multi-Messenger Nova: External Shocks, TeV Photons, and Neutrinos in the Next Eruption of T CrB}
\author[Aydi et al.]{\parbox{\textwidth} {E. Aydi\orcid{0000-0001-8525-3442}$^{1}$\thanks{E.~Aydi; E-mail: eaydi@ttu.edu},  P.~Craig$^{2}$, K.~V.~Sokolovsky\orcid{0000-0001-5991-6863}$^{1,5}$, L.~Chomiuk\orcid{0000-0002-8400-3705}$^{2}$, C.~C.~Cheung$^{3}$, L.~Izzo\orcid{0000-0001-969 5-8472}$^{4}$, J.~D.~Linford\orcid{0000-0002-3873-5497}$^{5}$,  B.~D.~Metzger\orcid{0000-0002-4670-7509}$^{6, 7}$, S. Mohamed$^{8}$, I. Molina$^{2}$, K.~Mukai$^{9, 10}$, K.~J.~Shen$^{11}$, and J.~L.~Sokoloski$^{6}$}
\vspace{0.4cm}\\
\parbox{\textwidth}{
$^{1}$Department of Physics and Astronomy, Texas Tech University, Lubbock, TX 79409, USA\\
$^{2}$Center for Data Intensive and Time Domain Astronomy, Department of Physics and Astronomy, Michigan State University, East Lansing, MI 48824, USA\\
$^{3}$Space Science Division, Naval Research Laboratory, Washington, DC 20375, USA
$^{4}$INAF-Osservatorio Astronomico di Capodimonte, Salita Moiariello 16, 80131, Napoli, Italy\\
$^{5}$National Radio Astronomy Observatory, P.O.\ Box O, Socorro, NM 87801, USA\\
$^{6}$Department of Physics and Columbia Astrophysics Laboratory, Columbia University, New York, NY 10027, USA\\
$^{7}$Center for Computational Astrophysics, Flatiron Institute, 162 5th Ave, New York, NY 10010, USA\\
$^{8}$Virginia Institute for Theoretical Astronomy, University of Virginia, Charlottesville, VA 22904, USA\\
$^{9}$CRESST and X-ray Astrophysics Laboratory, NASA/GSFC, Greenbelt, MD 20771, USA\\
$^{10}$Department of Physics, University of Maryland, Baltimore County, 1000 Hilltop Circle, Baltimore, MD 21250, USA\\
$^{11}$Department of Astronomy and Theoretical Astrophysics Center, University of California, Berkeley, CA 94720, US\\
}}  

\begin{document}
\label{firstpage}
\pagerange{\pageref{firstpage}--\pageref{lastpage}}
\maketitle

\begin{abstract}
T Coronae Borealis is the most compelling anticipated Galactic recurrent nova and one of the best opportunities to test whether nova eruptions can become genuine multi-messenger transients. Recent observational work on classical novae has shown that the GeV $\gamma$-ray luminosity correlates strongly with the differential velocity between interacting outflows, suggesting that shock power depends strongly on velocity contrast. We apply this empirical framework to T~CrB and compare it with the 2021 eruption of RS~Ophiuchi, the only nova securely detected at TeV energies. We argue that T~CrB may lie at the extreme high-$\Delta v$ end of the nova population and that its strongest shock should arise at the external interaction between the nova ejecta and the slow circumbinary medium. This makes the external shock the most likely site of any TeV emission, while the GeV signal may contain contributions from both internal, if present, and external shocks. Because T~CrB is substantially closer than RS~Oph, it is an especially favorable target for TeV detection. The neutrino case is promising but more uncertain because it depends sensitively on the density and structure of the circumbinary material. We also revisit the recurrence time of T~CrB. Using the observed scatter of better-sampled recurrent novae, we infer a representative population-scale recurrence of $T_{\rm rec}\simeq80\pm12$~yr. Taken together, its proximity and potentially extreme shock velocities make T CrB perhaps the best opportunity for a nova to become the first securely established multi-messenger nova. 
\end{abstract}

\begin{keywords}
stars: novae, cataclysmic variables --- shocks --- gamma rays --- neutrinos --- cosmic rays
\end{keywords}

\section{Introduction}

Classical and recurrent novae --- outbursts in systems with accreting white dwarfs powered by thermonuclear runaways --- are now recognized to commonly host energetic shocks that can contribute substantially to their multi-wavelength emission. \citep{Li_etal_2017_nature,Aydi_etal_2020a,Chomiuk_etal_2021}. This was established by the now routine detection of GeV gamma-rays from more than 20 Galactic novae by the Large Area Telescopes (LAT) on NASA's Fermi $\gamma$-ray satellite \citep{Abdo_etal_2010,Ackermann_etal_2014,Cheung_etal_2016,Franckowiak_etal_2018,Craig_etal_2026}. 
In classical novae, high-cadence optical spectroscopic studies have shown that eruptions commonly produce multiple distinct outflows: 
a slower flow is followed by a faster one, and their collision produces shocks capable of generating GeV $\gamma$-rays \citep{Aydi_etal_2020b,Craig_etal_2026,Aydi_etal_2026}. 
In at least one exceptionally well-observed event, V906 Car,
 the optical and $\gamma$-ray light curves 
showed a series of quasi-simultaneous flares, 
demonstrating that shocks can contribute a substantial fraction of the nova luminosity itself \citep{Aydi_etal_2020a}. 
These results have transformed the view of novae from purely thermal transients into laboratories for radiative shocks and particle acceleration.

The 2021 eruption of the recurrent nova RS Ophiuchi extended this picture to much higher energies. The detection of TeV $\gamma$-ray emission by H.E.S.S., MAGIC, and CTAO LST-1 demonstrated that nova shocks can accelerate particles to at least TeV energies in symbiotic systems, where fast nova ejecta interact with the dense wind of a red-giant companion \citep{HESS_2022,Acciari_etal_2022,2025A&A...695A.152A}. These detections resolved a long-standing theoretical question over whether nova shocks could accelerate particles to TeV energies \citep{Razzaque_etal_2010,2012PhRvD..86f3011S,2013A&A...551A..37M,Metzger_etal_2016,Zheng_etal_2022,DeSarkar_etal_2023,Zheng_etal_2024}, following a series of previous TeV non-detections from other novae \citep{2012ApJ...754...77A,2015arXiv150205853S,2022ApJ...940..141A}. The observed GeV--TeV emission from RS~Oph has also been used to estimate the corresponding high-energy neutrino emission and its detectability with current and future neutrino facilities \citep{Guetta_etal_2023}.
Subsequent modelling of the RS Oph high-energy data suggested that a single shock cannot simultaneously explain the full GeV--TeV phenomenology 
and favored multiple shock components, potentially including anisotropic external shocks and internal ejecta collisions \citep{Diesing_etal_2023}. 
Thus, although the existence of particle acceleration is no longer in doubt, the location of the dominant high-energy emission zone remains an open question. 

T Coronae Borealis (T CrB) is the natural next target. It is a symbiotic recurrent nova with historical eruptions in 1866 and 1946, with possible earlier eruptions in 1787 and 1217, implying a recurrence time-scale of roughly 80~yr \citep{Mclaughlin_1946,Schaefer_etal_2023,Starrfield_etal_2024}. In recent years, T CrB has already motivated dedicated neutrino forecasts \citep{Thwaites_Vandenbroucke_2025}, two-zone multi-messenger models involving both external shocks and magnetic reconnection \citep{Sarmah_etal_2025}, and detailed hydrodynamic and particle-acceleration calculations \citep{Orlando_etal_2025,Petruk_etal_2026}. These studies highlight the exceptional multi-messenger potential of the system, although their predictions are necessarily model-driven, i.e. sensitive to assumptions about the ejecta, circumbinary environment, and particle-acceleration physics.

In this paper we adopt a deliberately simpler and more observationally anchored approach. Our starting point is the empirical result of \citet{Craig_etal_2026}, who showed that the GeV luminosity of novae correlates strongly with the velocity differential between interacting flows. We then ask what this implies for T CrB if both internal and external shocks are present. Our main claim is that the largest velocity contrast in T CrB should occur at the external shock, between the nova ejecta and the slow circumbinary medium, and that this makes the external shock the most plausible origin of any TeV emission. In contrast, the GeV signal may be a mixture of internal (if present) and external shock contributions. If so, T CrB may represent one of the best opportunities for a nova to be detected in TeV photons and perhaps neutrinos, thereby becoming the first bona fide multi-messenger nova.

\section{Physical Motivation}

\subsection{Velocity contrast as a leading control parameter}
\label{sec_velocity_contrast}

A central recent observational result by \citet{Craig_etal_2026} shows that the GeV $\gamma$-ray luminosity of novae correlates with the velocity contrast between interacting outflows. In their Galactic nova sample, systems with faster outflows---and correspondingly larger velocity differences between the fast and slow components---show higher average GeV $\gamma$-ray luminosities and shorter $\gamma$-ray durations. This trend is qualitatively consistent with expectations for emission originating in shocks, since the kinetic power processed by a shock scales as
\begin{equation}
L_{\rm sh} \sim \frac{1}{2}\rho A(\Delta v)^3,
\end{equation}
where $\rho$ is the upstream density, $A$ is the shock area, and $\Delta v$ is the relative velocity across the shock. Thus, for otherwise comparable conditions, even a modest increase in velocity contrast can lead to a substantial increase in the available shock power.


In this paper we therefore adopt an empirically motivated hierarchy in which the differential velocity,
\begin{equation}
\Delta v = v_{\rm fast} - v_{\rm slow},
\end{equation}
provides the primary observational proxy for the strength of the shock. These velocities are inferred from the radial velocities of the spectroscopic components, particularly the absorption troughs of P~Cygni profiles, and therefore may differ from the true three-dimensional outflow velocities because of inclination and ejecta geometry. Nevertheless, in the absence of spatially resolved kinematic information, these spectroscopic velocities provide our best observational estimate of the relative speeds of the interacting flows. A larger velocity contrast may also favor acceleration to higher particle energies: in addition to increasing the available shock power, the maximum particle energy can depend strongly on shock velocity and on the spatial extent of the hot post-shock acceleration region, which can be substantially reduced by radiative cooling and turbulent mixing \citep{Diesing_Metzger_2026}.

Density is also a fundamental ingredient in determining the shock power and the efficiency of high-energy emission. Indeed, nova ejecta masses, and consequently their densities, can span several orders of magnitude, substantially more than the factor of a few spanned by the characteristic outflow velocities. Despite this large potential variation in density, \citet{Craig_etal_2026} find a significant correlation between GeV $\gamma$-ray luminosity and $\Delta v$, broadly consistent with the expected $L_{\rm sh}\propto \rho A(\Delta v)^3$ scaling. The scatter about this relation likely reflects, at least in part, differences in ejecta density, mass, and geometry. Importantly, however, faster novae---which generally have more rapidly expanding and therefore more dilute ejecta at a given time and ejecta mass---nevertheless tend to have the highest $\gamma$-ray luminosities, whereas slower novae, whose ejecta remain denser for longer, tend to be less luminous in $\gamma$-rays. This empirical behaviour suggests that the strong velocity dependence of the shock power can dominate over the opposing density trend. We therefore treat $\Delta v$ as a leading observational discriminator of shock strength, while density remains a key physical parameter governing the shock energetics, radiative efficiency, and the importance of hadronic processes such as proton--proton interactions and neutrino production.

This empirical picture is especially useful for T CrB because the system possibly hosts both internal and external shocks. Internal shocks arise from collisions between multiple ejecta components, as in classical novae \citep{Aydi_etal_2020b}. External shocks arise as the ejecta plow into the red-giant environment, as in symbiotic novae such as V407~Cyg \citep{Ackermann_etal_2014} and RS Oph \citep{HESS_2022,Acciari_etal_2022}. The key question is therefore not whether shocks exist, but which shocks should dominate the highest-energy output. We emphasize, however, that the empirical relation established by \citet{Craig_etal_2026} applies primarily to GeV emission associated with internal shocks in classical novae. Its extension to external shocks in symbiotic systems, and particularly to the production of TeV emission, has not yet been established observationally. We therefore treat the application of this framework beyond the regime explored by \citet{Craig_etal_2026} as a working hypothesis rather than a demonstrated scaling. Density and geometry may also significantly affect the resulting high-energy emission, particularly the efficiency of hadronic interactions. These considerations are especially relevant for T~CrB, where the density and geometry of the circumbinary environment remain important uncertainties.

\subsection{RS Oph as the benchmark}

RS Oph provides the benchmark because it is the only nova securely detected at TeV energies \citep{HESS_2022,Acciari_etal_2022}. In the standard symbiotic-nova picture, the fast ejecta interact with the red-giant wind and drive a powerful external shock \citep{Tatischeff_Hernanz_2007,Tatischeff_Hernanz_2023}. Representative values used in the recent literature are a fast flow of order 
$v_{\rm ej,RSOph} \approx 3700~{\rm km~s^{-1}}$ and a slow, circumbinary flow of order $v_{\rm CBM,RSOph} \approx 50~{\rm km~s^{-1}}$ \citep{Diesing_etal_2023},
which imply $\Delta v_{\rm ext,RSOph} \approx 3650~{\rm km~s^{-1}}$ (these values are measured from the absorption troughs of the Balmer P Cygni lines during the first days of the eruption). By contrast, a representative internal-shock contrast for RS Oph is smaller, for example between $\sim 2700$ and $\sim 3700~{\rm km~s^{-1}}$, which implies $\Delta v_{\rm int,RSOph} \approx 1000~{\rm km~s^{-1}}$. Thus, even before density is considered, the external shock in RS Oph naturally provides the largest available velocity difference and potentially the greatest shock power. 

At the same time, \citet{Diesing_etal_2023} argued that the GeV and TeV data from RS Oph cannot be explained by a single shock and favored a multi-shock picture. Importantly, this result does not undermine the role of the external shock. Rather, it suggests that the high-energy emission may arise from several distinct shock zones, including external shocks propagating through an anisotropic circumbinary medium---for example, interacting with denser material concentrated toward the binary orbital plane and lower-density material toward the poles---as well as possibly from internal collisions between distinct components of the nova ejecta. 

In that context, the highest-energy photons may preferentially originate from the shock with the largest velocity contrast. One possibility is that a substantial fraction of the GeV emission arises from internal shocks between distinct ejecta components, while the TeV emission is dominated by the external shock driven into the red-giant environment. Because these shocks form and evolve on different spatial and temporal scales, such a configuration could contribute to the several-day delay between the peaks of the GeV and TeV light curves in RS~Oph \citep{HESS_2022,Diesing_etal_2023}. This interpretation is not unique: an anisotropic circumbinary medium could instead produce multiple external shocks, with the denser material concentrated toward the orbital plane contributing predominantly at GeV energies and a faster shock propagating through the lower-density, more nearly polar wind producing the TeV emission \citep{Diesing_etal_2023}.

\subsection{Historical shock kinematics of T CrB}

For T CrB, historical spectroscopy provides the key observational clue. \citet{Mclaughlin_1946} reported a very fast absorption component of around $4000~{\rm km~s^{-1}}$ during the early days of the 1946 eruption. Recent hydrodynamic modelling of T CrB has adopted a red-giant wind speed of $\approx 10~{\rm km~s^{-1}}$ as part of its baseline circumbinary-medium model \citep{Orlando_etal_2025}. Taken together, these imply an external-shock contrast of $\Delta v_{\rm ext,TCrB} \approx 4000 - 10 \approx 3990~{\rm km~s^{-1}}$. This is somewhat larger than the representative external-shock velocity contrast inferred for RS~Oph and significantly larger than the differential velocities typically measured between the `slow' and `fast' flows in classical novae, placing T~CrB on the extreme end of the nova population in terms of the differential velocity between interacting ejecta.

T CrB might also host strong internal shocks. Using the historical values discussed in \citet{Mclaughlin_1946}, a representative internal collision in T CrB may occur between flows at $\sim 1800$ and $\sim 4000~{\rm km~s^{-1}}$ (measured from the absorption troughs of the Balmer lines), which gives
$\Delta v_{\rm int,TCrB} \approx 2200~{\rm km~s^{-1}}$. 
This is also larger than the representative internal-shock contrast usually discussed for RS Oph. Therefore, T CrB may host stronger internal shocks than RS Oph, but its strongest shock should still be the external one. These measurements motivate our central interpretation. Internal shocks can plausibly contribute to the GeV signal, perhaps substantially, but the TeV emission could preferentially trace the external shock because that is where $\Delta v$ is largest. We operate under the \emph{assumption} that the luminosity in both GeV and TeV scales with velocities in the same way (i.e. $L_{\mathrm{Gev}} \propto dv^3$ and $L_{\mathrm{Tev}} \propto dv^3$).

\begin{figure*}
\begin{center}
  \includegraphics[width=1.0\textwidth]{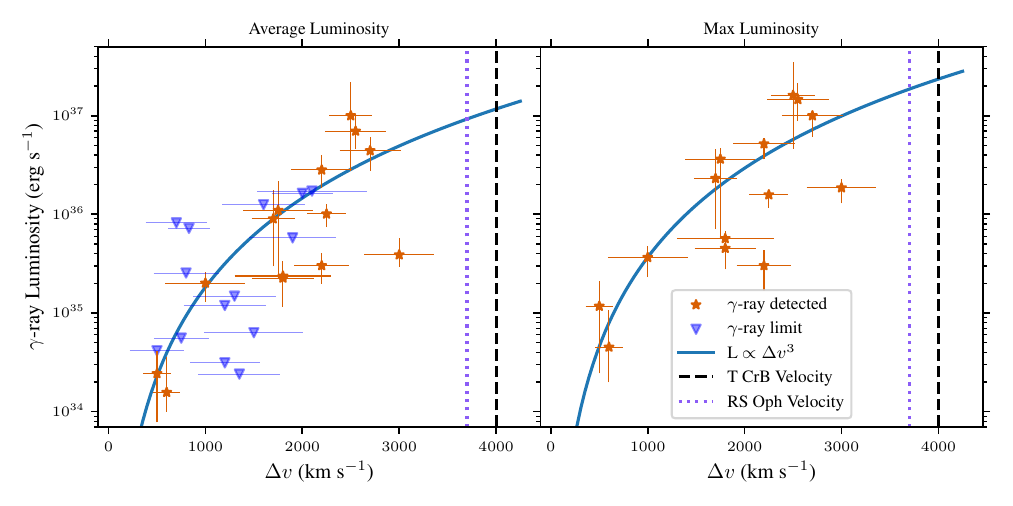}
\caption{The $\gamma$-ray luminosity of the classical novae from \citet{Craig_etal_2026} plotted against the relative velocity between the slow and fast outflows ($v_2 - v_1 = \Delta v$). We also highlight the differential velocities of RS~Oph (dotted vertical line) and T~CrB (dashed vertical line) in order to compare with the sample of \citet{Craig_etal_2026}. The left panel displays the average $\gamma$-ray luminosity over the duration of \emph{Fermi} detection, while the right panel displays the maximum $\gamma$-ray luminosity. The orange stars show novae with \emph{Fermi}-detected $\gamma$-rays, while the blue triangles represent upper limits for novae that lack a $\gamma$-ray detection. The blue solid line shows a fit to the $\gamma$-ray detections assuming the scaling $L_\gamma=a(\Delta v)^3$, where $a$ is the fitted normalization.} 
\label{Fig:Lgamma_dv}
\end{center}
\end{figure*}

In Figure~\ref{Fig:Lgamma_dv} we compare the external-shock differential velocities inferred for RS~Oph and T~CrB with the differential velocities measured for internal shocks in classical novae by \citet{Craig_etal_2026}. We emphasize that internal and external shocks represent different physical environments and therefore do not constitute a direct one-to-one comparison. Rather, the purpose of this comparison is to place the velocity contrasts expected in the symbiotic recurrent novae within the broader range observed in nova shocks. RS~Oph, and particularly T~CrB, lie toward the extreme high-$\Delta v$ end of this parameter space, suggesting the potential for correspondingly powerful shocks. However, because the density, geometry, and other physical conditions governing internal shocks in classical novae can differ substantially from those of external shocks in symbiotic systems, we do not use this comparison to predict a $\gamma$-ray luminosity for T~CrB. Instead, Figure~\ref{Fig:Lgamma_dv} is intended primarily to demonstrate how the expected external-shock velocity contrast in T~CrB compares with the differential velocities inferred for the classical-nova population.

\section{An Observationally Anchored prediction}

\subsection{Why the TeV emission is likely external}

If the empirical relation identified by \citet{Craig_etal_2026} is taken seriously, then the strongest high-energy output should arise where the velocity contrast is largest. In RS Oph, that naturally points to the external shock, and indeed TeV emission was detected in that system \citep{HESS_2022}. In T CrB, the corresponding external-shock, velocity-contrast may be even larger (see previous sections). This suggests that T~CrB could be at least as promising a target as RS~Oph for TeV detection, although differences in circumbinary density remain an important uncertainty (see sections~\ref{sec_velocity_contrast} and~\ref{sec_density}). 

This line of reasoning also explains why TeV emission has not been seen from ordinary classical novae. In those systems, the main shocks are believed to be internal collisions between slow and fast ejecta components with typical velocity differences of order a few $10^2$ to a few $10^3~{\rm km~s^{-1}}$ \citep{Aydi_etal_2020b,Craig_etal_2026}. In symbiotic recurrent novae such as RS Oph and T CrB, the relevant external shock involves ejecta at several thousand ${\rm km~s^{-1}}$ impacting a circumbinary medium moving at only $\sim 10$--$50~{\rm km~s^{-1}}$, producing a larger $\Delta v$ and, in turn, a stronger accelerator (Figure~\ref{Fig:Lgamma_dv}). We suggest that this larger external velocity contrast may be an important reason why RS~Oph was detected at TeV energies while classical novae have not been. However, it is worth noting that the absence of TeV detection from classical novae is strongly affected by selection effects: current imaging atmospheric Cherenkov telescopes have relatively narrow fields of view and limited duty cycles, and only a small number of nova eruptions have received sufficiently prompt and sensitive TeV follow-up. The absence of TeV detections from classical novae therefore does not necessarily imply that such systems are intrinsically incapable of producing TeV emission. Rather, their generally smaller internal velocity contrasts may make detectable TeV emission less favorable than in symbiotic novae such as RS~Oph and T~CrB, where fast ejecta interact with a much slower circumbinary medium.

The RS Oph literature supports the interpretation that TeV emission is more favorable in symbiotic systems where multiple shocks could be present, even if it does not uniquely identify a single zone. \citet{Diesing_etal_2023} showed that the GeV--TeV phenomenology likely requires multiple shocks, but they did not establish that the TeV signal originated purely in internal or external shocks. Instead, they left open a picture involving several shocks, including distinct external shocks in an anisotropic circumbinary environment. We therefore consider it both conservative and physically motivated to associate the dominant TeV component with the external shock.

\subsection{The role of density}
\label{sec_density}

The strongest caveat to the above argument is the density of the circumbinary target material. \citet{Orlando_etal_2025} inferred a relatively low red-giant wind mass-loss rate for T CrB ($\dot{M}_{\rm RG} \approx 4 \times 10^{-9}~M_\odot~{\rm yr^{-1}}$), 
for a $10~{\rm km~s^{-1}}$ wind, and concluded that the circumbinary medium of T CrB is less dense than in other symbiotic novae (see also \citealt{Linford_etal_2019}). If this estimate is correct, then the hadronic interaction efficiency of the external shock may be lower than in RS Oph, which would weaken both the neutrino signal and the hadronic component of the $\gamma$-ray emission.

However, two points are important here. First, this low-density estimate is not yet beyond debate. Second, even if the density is somewhat lower, the observational evidence from classical novae suggests that velocity contrast is the leading parameter in setting the shock power (see section~\ref{sec_velocity_contrast}). In this interpretation, velocity contrast provides a strong observational proxy for the available shock power, while density and geometry regulate both the shock energetics and the efficiency with which that power emerges as high-energy radiation. This is exactly why T CrB remains such a promising source even under relatively conservative environmental assumptions.

There is also an independent argument for caution in treating the low wind-density estimate as settled. The recurrence time-scale of T CrB requires a high mass-accretion rate on~to a massive white dwarf. Recent modelling finds that reproducing the $\sim 80$~yr recurrence demands $\dot{M}_{\rm acc} \sim 10^{-8} - 10^{-7}~M_\odot~{\rm yr^{-1}}$ for $M_{\rm WD}\sim 1.30$--$1.38~M_\odot$ \citep{Toala_etal_2024,Jose_Hernanz_2025}. This does not imply that the freely expanding red-giant wind must be equally dense, since transfer may be relatively efficient if the giant fills or nearly fills its Roche lobe. Nevertheless, it reinforces the idea that the relationship between donor mass loss, accretion flow, and circumbinary density is not yet uniquely determined.

\subsection{Distance and observed flux}

Distance provides a further advantage for T CrB. Recent T CrB multi-messenger forecasts typically adopt a distance near $0.8$~kpc, whereas RS Oph is usually placed between $1.4$ and $2.5$~kpc \citep{Thwaites_Vandenbroucke_2025,Sarmah_etal_2025}. At fixed luminosity, this implies an observed-flux enhancement for T CrB by approximately

$$
\left(\frac{d_{\rm RSOph}}{d_{\rm TCrB}}\right)^2
\simeq 3.1\mbox{--}9.8.
$$

Thus, even if T CrB and RS Oph had comparable intrinsic high-energy luminosities, T CrB would be expected to appear approximately $3$--$10$ times brighter at Earth purely as a consequence of its smaller distance.

Combined with the slightly larger external-shock velocity contrast in T CrB, this makes a strong case that T CrB should be at least as detectable as RS Oph in TeV $\gamma$-rays. A caveat is that the circumstellar environment of RS Oph is expected to be denser than that of T CrB (see Section~\ref{sec_density}). Nevertheless, our comparison suggests that the differential velocity between the interacting ejecta and/or external medium, may play a particularly important role in determining the strength of the shocks and the resulting high-energy emission.

\subsection{Prospects for TeV Photons, Neutrinos, and Cosmic Rays}

\subsubsection{TeV $\gamma-ray$ emission}

We expect T~CrB to be detected in GeV $\gamma$-rays if its next eruption resembles other novae in which strong shocks produce high-energy emission. The more interesting question is whether it will also be visible at TeV energies. We argue that T~CrB is an exceptionally promising TeV candidate because its external shock likely combines a larger velocity contrast than RS Oph with a significantly smaller distance.

This conclusion is broadly consistent with recent detailed modelling. The two-zone T CrB study of \citet{Sarmah_etal_2025} found that the external shock readily produces detectable $\gamma$-ray emission, while the more detailed hydrodynamic plus particle-acceleration calculations of \citet{Petruk_etal_2026} found that TeV emission should persist for roughly a week and that the early high-energy signal is dominated by the ejecta-driven forward shock. Our argument differs in being observationally motivated rather than simulation-based, but the physical conclusion is similar: the external shock is the best candidate for the highest-energy photons.

\subsubsection{Neutrino emission}

The prospects for detecting high-energy neutrinos from T CrB are more uncertain than those for TeV $\gamma$-rays. In hadronic scenarios, neutrinos are produced when shock-accelerated protons interact with the surrounding material, so the expected neutrino yield depends sensitively on the target density and column encountered by the accelerated particles, as well as on the energy transferred to relativistic protons \citep{Razzaque_etal_2010,DeSarkar_etal_2023}. This dependence is particularly important for T CrB because the density and geometry of its circumstellar environment remain uncertain. The dedicated IceCube T CrB study of \citet{Thwaites_Vandenbroucke_2025} nevertheless concluded that T CrB should be a more promising neutrino target than RS Oph because it is closer, was optically brighter during its historical eruptions, and is located at a more favorable declination for IceCube. However, that study largely rescales expectations from RS Oph rather than predicting the neutrino emission from a detailed physical model of the T CrB environment. The two-zone analysis of \citet{Sarmah_etal_2025} found that the external-shock neutrino signal is generally modest, with detectability only in more optimistic regions of parameter space, while \citet{Petruk_etal_2026} likewise found that neutrino detection is primarily feasible in models combining high explosion energy with high ambient density. More broadly, previous studies illustrate both the potential and the difficulty of detecting high-energy neutrinos from novae. Using the observed GeV--TeV emission from the 2021 eruption of RS~Oph as a benchmark, \citet{Guetta_etal_2023} estimated the associated neutrino emission and examined its detectability with present and future neutrino facilities. They found that the expected signal from RS~Oph itself was below the sensitivity of current detectors, but argued that next-generation facilities such as IceCube-Gen2 could detect neutrinos from Galactic novae at a rate of order once per decade, potentially increasing to once every few years depending on the Galactic nova rate. For T~CrB specifically, \citet{Wang_etal_2025} predict that its high-energy hadronic neutrino flux would remain below the sensitivity of the current-generation IceCube detector.

Our interpretation is therefore cautious. T CrB is plausibly one of the most promising nova neutrino targets to date, but a neutrino detection is not guaranteed. The neutrino yield will depend on whether the large shock power associated with the external $\Delta v$ occurs in a sufficiently dense environment to process and accelerate a substantial population of particles and to provide an efficient target for subsequent hadronic interactions. In this sense, T CrB provides an important test case for the interplay between shock energetics, particle acceleration, and the density of the circumstellar environment. A TeV detection without a corresponding neutrino detection would place useful constraints on the hadronic efficiency and physical conditions of the emitting region, although it would not by itself distinguish between a low effective target density, inefficient hadronic acceleration, or other differences in the underlying particle population. A joint TeV-plus-neutrino detection would provide substantially stronger evidence that nova shocks are efficient hadronic accelerators.

IceCube is the natural flagship instrument because T CrB lies in the northern sky, but coordinated observations with KM3NeT and lower-energy IceCube event selections will also be important \citep{Thwaites_Vandenbroucke_2025}. Simultaneous GeV and TeV monitoring will be essential for interpreting any neutrino signal.

\subsubsection{Cosmic rays}

T~CrB is very likely to accelerate ions to relativistic energies if it follows the broad phenomenology established by RS~Oph \citep{HESS_2022}. In that sense, it should be regarded as a likely cosmic-ray accelerator. The relative importance of escaping cosmic rays and neutral secondaries is governed by the competition between particle escape and interaction: accelerated particles that escape the nova environment contribute to the cosmic-ray population, whereas those undergoing hadronic interactions before escape can produce $\gamma$-rays and neutrinos. The density and structure of the surrounding material therefore play an important role in determining this balance. However, direct detection of cosmic rays from T~CrB is probably less likely because charged particles are deflected by Galactic magnetic fields. The most useful observational signatures are therefore the neutral secondaries, $\gamma$-rays and neutrinos. A combined TeV-plus-neutrino detection from T~CrB would provide the strongest practical evidence to date that nova shocks accelerate hadrons and contribute to the non-thermal particle population.

\section{Discussion}
\subsection{T CrB in the literature}

T CrB has recently become the focus of a rapidly expanding theoretical and observational literature, reflecting its status as one of the most compelling anticipated Galactic recurrent novae and an unusually favorable opportunity to test whether nova eruptions can produce detectable high-energy and multi-messenger emission. A common theme across this work is that the next eruption is expected to drive strong shocks into a complex symbiotic environment, although substantial uncertainty remains regarding the geometry of those shocks, the efficiency of particle acceleration, and the relative importance of the different interaction regions.

Several recent studies have approached this problem from complementary directions. Observationally motivated work has emphasized detectability, particularly in neutrinos, by scaling from RS~Oph and from the historical optical behaviour of T~CrB \citep{Thwaites_Vandenbroucke_2025}. More detailed high-energy calculations have explored multiple possible particle-acceleration sites. \citet{Sarmah_etal_2025}, for example, considered both an external shock formed by the interaction of the nova ejecta with the red-giant environment and a separate magnetic-reconnection scenario close to the white dwarf. The latter should be regarded as a more model-dependent possibility, since there is currently no clear observational evidence that the white dwarf in T~CrB possesses a sufficiently strong magnetic field for reconnection near the WD to dominate the high-energy particle budget.

Hydrodynamic calculations provide a firmer physical basis for interpreting the high-energy emission in terms of shocks. The three-dimensional simulations of \citet{Orlando_etal_2025} show that the interaction of the nova ejecta with the T~CrB circumbinary medium should be strongly structured by the red-giant wind, an equatorial density enhancement, the accretion disc, and the companion star. These calculations also indicate that the circumbinary density is lower than in RS~Oph, with corresponding consequences for the shock evolution and high-energy emission. Building directly on this hydrodynamic framework, \citet{Petruk_etal_2026} model particle acceleration through diffusive shock acceleration at the forward shock. Their calculations are spatially multi-zone in the sense that the shock properties, downstream flow, particle spectra, and emission vary throughout the evolving interaction region; however, the underlying particle-acceleration mechanism remains shock acceleration. They predict both hadronic $\gamma$-ray emission from proton--proton interactions and leptonic emission from inverse-Compton scattering, with the resulting fluxes depending sensitively on explosion energy, ambient density, geometry, and post-shock flow structure.

At the same time, nova-evolution calculations reinforce the expectation that T~CrB contains a massive white dwarf undergoing relatively high mass transfer. Hydrodynamic nova models find that reproducing an $\sim80$ yr recurrence time requires a WD mass of approximately $1.30$--$1.38\,M_\odot$ for plausible accretion rates and compositions, although the resulting explosion properties depend sensitively on the WD luminosity, accretion rate, metallicity, and composition \citep{Jose_Hernanz_2025}. These calculations support the expectation of an energetic nova eruption but also emphasize that the explosion energetics cannot be inferred from WD mass or recurrence time alone.

Complementary predictions have also been made for the thermonuclear eruption itself and its nucleosynthetic signatures. \citet{Starrfield_etal_2025} performed one-dimensional hydrodynamic calculations of thermonuclear runaways on a massive $\sim1.35,M_\odot$ WD, exploring both CO and ONe compositions and accretion rates chosen to produce recurrence times of approximately 80~yr. These models predict a luminous eruption with strong early X-ray and UV emission and substantial retention of accreted material by the WD. Building on these calculations, \citet{Wallace_etal_2025} investigated the predicted elemental and isotopic yields while explicitly propagating uncertainties in the thermonuclear reaction rates. They showed that measurements of the ejecta composition, particularly of elements such as S, Ne, Si, and P, could provide important constraints on whether the underlying WD is CO or ONe. These studies emphasize that the next eruption of T~CrB will provide constraints not only on shock acceleration and the circumbinary environment, but also on the thermonuclear runaway, WD composition, and long-term evolution of the system.

Taken together, the recent literature therefore favors a picture in which shock acceleration provides the most observationally and physically motivated route to high-energy emission from T~CrB. The principal uncertainty is not whether a fundamentally different mechanism is required, but rather which shock regions dominate the particle acceleration and radiation, and how their efficiency is regulated by the velocity structure, density, and geometry of the circumbinary medium. Magnetic reconnection close to the WD has been proposed as an additional possible acceleration channel \citep{Sarmah_etal_2025}, particularly for neutrino production, but this scenario currently rests on stronger assumptions about the WD magnetic environment than the external-shock interpretation. The existing literature has therefore established T~CrB as a plausible high-energy source while leaving open the observationally important question of which shock properties will determine the TeV output.

\subsection{Interpreting T CrB in the context of recent work}
The main strength of the picture proposed here is that it is rooted in observation. We do not attempt a full hydrodynamic reconstruction of the T CrB environment. Instead, we ask which shock in T CrB should be strongest in the empirical sense established by \citet{Craig_etal_2026}. The answer is the external shock, because that is where the velocity contrast is largest. This is particularly important in the case of T~CrB, which may represent an extreme case among novae: the differential velocity between its fast ejecta and slow circumbinary environment appears to lie at the upper end of the observed nova population. If differential velocity is indeed the leading control parameter for shock power, then this makes T~CrB an especially compelling candidate for TeV emission and possibly other high-energy secondaries.

This framework also clarifies how to interpret the growing recent T CrB literature. The IceCube strategy paper of \citet{Thwaites_Vandenbroucke_2025} is useful as an observational benchmark, as it also adopt RS Oph-based scaling arguments. The multi-messenger study of \citet{Sarmah_etal_2025} broadens the allowed parameter space by including both external shocks and magnetic reconnection near the white dwarf, but its more optimistic neutrino predictions come from the reconnection channel rather than from the external shock. The hydrodynamic and particle-acceleration studies of \citet{Orlando_etal_2025} and \citet{Petruk_etal_2026} provide a more detailed treatment of the circumbinary environment, but necessarily depend on assumed ejecta and environmental parameters, including the explosion energy, ejecta mass, ejecta velocity structure, and circumbinary density distribution. In particular, these simulations were not calibrated to reproduce the historical $\sim4000~{\rm km~s^{-1}}$ absorption component observed during the 1946 eruption. Our argument complements these calculations by using that historical velocity measurement directly and isolating a particularly useful observational lever arm: the velocity contrast.

The largest open issue is the density of the T CrB circumbinary medium. If it is substantially lower than in RS Oph, the neutrino yield may be modest. But even in that case, T CrB remains one of the best available targets for studying hadronic acceleration in novae because its large velocity contrast and proximity make it extremely favorable for $\gamma$-ray detection. Conversely, if later observations show that the red-giant mass loss or effective target density is larger than currently assumed, then the case for neutrino detection strengthens considerably.

A second caveat is that the next eruption need not exactly repeat the 1946 event. Nevertheless, recurrent novae are sufficiently repeatable that the historical kinematics provide a meaningful guide, and the presence of a symbiotic environment ensures that an external shock will exist even if the exact flow speeds differ somewhat from those inferred historically.

\subsection{Recurrence time and uncertainty in the next eruption}

Recurrent novae are not strictly periodic systems. The recurrence time-scale is determined by several coupled properties of the binary and the white dwarf (WD), including the WD mass and thermal state, the mass-transfer and accretion rates, the chemical composition of the accreted material, and the ignition conditions at the base of the envelope. Multicycle nova calculations have demonstrated strong dependences of nova properties and recurrence times on $M_{\rm WD}$, the WD core temperature, and $\dot{M}$ \citep{Starrfield_etal_1985,Hachisu_Kato_2001,Yaron_etal_2005,Anupama_2013,Shara_etal_2018_June}. The composition of the accreted material can also modify the conditions for thermonuclear ignition and the amount of material required to trigger an eruption \citep{Shen_Bildsten_2009ApJ}, while calculations spanning different metallicities and mixing prescriptions show corresponding changes in the ignition mass and recurrence period \citep{Chen_etal_2019_Dec}. The recurrence time should therefore not be regarded as a clock controlled by a single parameter \citep{Henze_etal_2018}. Variations in the physical conditions of the system between successive cycles can modify the interval between eruptions, such that a recurrent nova is more appropriately characterized by both a typical recurrence interval, $T_{\rm rec}$, and an intrinsic cycle-to-cycle dispersion, $\sigma_{\rm rec}$. In the specific case of T~CrB, \citet{Luna_etal_2026} also emphasized that the accretion rate has varied substantially, providing an additional reason not to expect its eruptions to follow a strictly periodic clock (see also, \citealt{Merc_etal_2025}).

To investigate empirically whether this dispersion depends on recurrence time, we compiled the observed eruption histories of Galactic recurrent novae together with the two short-recurrence systems M31N~2008-12a and M31N~2017-01e. We emphasize that this is a heterogeneous population. Recurrent novae span different binary configurations, including systems with main-sequence, subgiant, and red-giant donors, and consequently can have substantially different mass-transfer rates and circumbinary environments. More generally, the recurrence time is determined by several physical parameters, including the white-dwarf mass and thermal state, the accretion rate and its history, and the composition and ignition conditions of the accreted material. The dispersion measured across this population should therefore be regarded as an empirical estimate of the characteristic cycle-to-cycle variability among recurrent novae, rather than as evidence that all systems share the same underlying recurrence physics.

Only directly observed eruptions are included in the primary sample; proposed missed eruptions are not inserted into the eruption sequences. For systems with sufficiently well-sampled modern histories, we calculate the mean and median inter-eruption intervals, the sample standard deviation $s_T$, and the Median Absolute Deviation (MAD), expressed here as a scaled robust estimator of the dispersion,

$$
\sigma_{\rm MAD}=1.4826\,{\rm median}\left|\Delta T_i-{\rm median}(\Delta T)\right|.
$$

The MAD-based estimator is less sensitive to outliers than the standard deviation and is therefore useful here because a small number of anomalously long or short recurrence intervals can substantially increase the ordinary standard deviation, $s_T$, and potentially bias the inferred characteristic recurrence-time dispersion. M31N~2008-12a provides the best-sampled example. Its observed eruption sequence has a mean recurrence time of approximately $0.996$ yr and a standard deviation of $\simeq0.11$ yr ($\simeq40$ d), consistent with the recurrence statistics reported from its multi-epoch monitoring \citep{Basu_etal_2024}. M31N~2017-01e has a longer mean recurrence time-scale of $\simeq2.5$ yr but a comparable absolute dispersion; its updated eruption ephemeris gives a mean recurrence period of $924.0\pm7.0$ d \citep{Shafter_etal_2024}. At longer recurrence times, the observed modern sequences of U~Sco and RS~Oph display variations of order years between successive eruptions, while T~Pyx has shown substantially larger changes in its recurrence intervals. The eruption histories and statistics derived uniformly from these intervals are summarized in Table~\ref{tab:rn_recurrence}.

Although the number of well-sampled recurrent novae is small and the population is heterogeneous, the observed systems show a tendency for the absolute recurrence-time dispersion to increase toward longer recurrence periods. This trend should not be interpreted as implying that $T_{\rm rec}$ uniquely determines $\sigma_{\rm rec}$. The physical parameters governing thermonuclear ignition differ substantially between systems, and the recurrence histories themselves can evolve. Moreover, the present sample provides no motivation for an exponential extrapolation of the recurrence uncertainty. We therefore adopt a deliberately simple phenomenological description based on the fractional recurrence dispersion,

$$
f_{\rm rec}\equiv\frac{\sigma_{\rm rec}}{T_{\rm rec}}.
$$

For the five systems in our relatively well-sampled statistical subset, the median fractional dispersion calculated using the ordinary standard deviation is $f_{\rm rec}\simeq0.16$. Replacing the standard deviation with the more robust MAD-based estimator gives a very similar characteristic value, $f_{\rm rec}\simeq0.15$. The persistence of a positive relation between absolute dispersion and recurrence time-scale when using a robust statistic suggests that the trend is not solely produced by a few anomalously long recurrence intervals. Nevertheless, the substantial system-to-system scatter and the small size of the available sample preclude determination of a precise universal scaling relation. We therefore adopt $f_{\rm pop}\simeq0.15$ as a representative population-scale fractional dispersion rather than as a universal property of recurrent novae.

This empirical population relation provides one possible estimate of the uncertainty associated with the long recurrence time of T~CrB. If no additional information specific to T~CrB is used, and the system is treated as a member of the recurrent-nova population with $T_{\rm rec}\simeq80$ yr, the expected population-scale dispersion is

$$
\sigma_{\rm pop}\simeq f_{\rm pop}T_{\rm rec}
\simeq0.15\times80~{\rm yr}
\simeq12~{\rm yr}.
$$

In this population-dominated interpretation, the recurrence time-scale of T~CrB can therefore be characterized approximately as

$$
T_{\rm rec}\simeq80\pm12~{\rm yr}.
$$

The quoted width should not be interpreted as a formal measurement uncertainty on the recurrence period of T~CrB. Rather, it represents the characteristic cycle-to-cycle dispersion obtained by applying the fractional scatter of the better-sampled recurrent-nova population to a system with an $\simeq80$ yr recurrence time. Relative to the securely observed 1946 eruption, this population-based estimate places the nominal next-eruption epoch near 2026, but with a characteristic uncertainty of approximately a decade. This result is broadly consistent with the simpler estimate presented by \citet{Luna_etal_2026}, who used the observed irregularity of the eruption intervals of U~Sco and RS~Oph to argue that extrapolating an $\sim80$~yr recurrence time for T~CrB should carry an uncertainty of order a decade. Our analysis extends this argument by considering a broader recurrent-nova sample and estimating the characteristic fractional recurrence-time dispersion empirically.

A substantially narrower estimate is obtained if significant weight is instead assigned to the proposed historical eruptions of T~CrB. In addition to the securely observed eruptions of 1866 and 1946, an eruption near the end of 1787 has been proposed from historical observations \citep{Schaefer_etal_2023}. If this identification is accepted, the resulting two consecutive recurrence intervals are approximately $78.39$ and $79.75$ yr, giving ${T}_{\rm rec}=79.07~{\rm yr}$ and $s_T=0.96~{\rm yr}.$

An additional eruption in AD~1217 has also been proposed \citep{Schaefer_etal_2023}. Connecting this event to the later eruption record requires an additional assumption about the number of nova cycles that occurred during the $\simeq570$~yr interval between 1217 and 1787. If this interval is divided into seven cycles, as adopted by \citet{Schaefer_etal_2023} and \citet{Pei_etal_2026}, the corresponding mean interval is $81.44$~yr. Together with the subsequent intervals of $78.39$ and $79.75$~yr, this gives ${T}_{\rm rec}=79.86$~yr and $s_T=1.53$~yr. However, the choice of seven cycles is not uniquely determined by the historical record itself. The neighbouring integer aliases are also plausible a priori: six cycles would imply an average interval of $\simeq95.0$~yr between 1217 and 1787, whereas eight cycles would imply $\simeq71.3$~yr. Both values lie within, or close to, the broad range suggested by the population-based estimate above. Moreover, selecting seven cycles necessarily minimizes the mismatch with the $\sim80$~yr recurrence inferred from the more recent eruptions, so the resulting small dispersion should not be regarded as an independent demonstration that T~CrB has maintained a highly regular recurrence clock. The narrow history-dominated estimate therefore depends both on accepting the proposed 1217 and 1787 identifications and on adopting the seven-cycle interpretation of the long historical gap. Thus, if the proposed historical identifications are accepted and given substantial statistical weight, and the 1217--1787 interval is assumed to contain seven recurrence cycles, T~CrB appears considerably more regular than would be inferred by treating it as a generic member of the recurrent-nova population. Under this history-dominated interpretation, the resulting recurrence pattern has a characteristic dispersion of only $\sim1$--$2$~yr and implies a nominal next-eruption epoch close to 2025--2026.

The two estimates are not necessarily contradictory, but instead correspond to different assumptions about which information should dominate the recurrence prediction. If little weight is assigned to the proposed historical eruptions, the recurrent-nova population provides the principal empirical constraint and gives a broad characteristic recurrence of approximately $80\pm12$ yr. Conversely, if substantial weight is assigned to the proposed historical T~CrB eruption record \emph{and} the 1217--1787 interval is taken to contain seven recurrence cycles, the resulting sequence is consistent with a much narrower recurrence time-scale, with a dispersion of only $\sim1$--$2$~yr.

These two cases can be summarized as: (1) $T_{\rm rec}=80\pm12~{\rm yr}$ (population dominated), and (2) $T_{\rm rec}\simeq79\mbox{--}80~{\rm yr}$ with $\sigma_{\rm rec}\simeq1\mbox{--}2~{\rm yr}$ (history dominated).

Finally, these symmetric recurrence intervals describe the distributions that would have been inferred before reaching the nominal expected eruption epoch. Once that epoch is reached without an eruption, the statistical problem changes: the absence of an eruption itself provides additional information, and the recurrence distribution must be conditioned on survival to the current epoch. This distinction is already relevant for T~CrB. A recent analysis using the three effective historical recurrence intervals finds $\overline{T}_{\rm rec}=79.86$ yr and $s_T=1.53$ yr, but, after explicitly conditioning on the absence of an eruption through 2026 July 11, obtains a modified probability distribution for the subsequent eruption epoch \citep{Pei_etal_2026}. Such probabilities should be regarded as empirical recurrence indicators rather than physical eruption probabilities. Thus, the $80\pm12$ yr and $\simeq79$--$80$ yr with $\sim1$--$2$ yr dispersion cases are best interpreted as two alternative recurrence priors whose relative importance depends on the weight assigned to the recurrent-nova population and to the proposed historical eruptions of T~CrB.

\section{Conclusions}

We have argued that T~CrB is plausibly the best candidate yet to become the first multi-messenger nova. Our main conclusions are as follows.

\begin{itemize}

\item Recent observations of the nova population show that the GeV $\gamma$-ray luminosity correlates with the differential velocity between interacting flows \citep{Craig_etal_2026}. This suggests that shock kinematics, and particularly $\Delta v$, provide an important observational predictor of the high-energy output, while density and geometry remain important in setting the shock power and radiative and hadronic efficiencies.

\item In T~CrB, the largest velocity contrast should occur at the external shock between the fast ejecta and the slow circumbinary medium. Historical and recent values imply $\Delta v_{\rm ext,TCrB}\approx4000~{\rm km~s^{-1}}$, somewhat larger than the representative RS~Oph value of $\approx3600~{\rm km~s^{-1}}$, placing T~CrB toward the extreme high-$\Delta v$ end of the nova population.

\item Internal shocks in T~CrB may also be present and could contribute substantially to the GeV emission. However, we argue that the external shock is the most plausible site of the highest-energy emission because it provides the largest velocity contrast and therefore potentially the greatest available shock power.

\item T~CrB is substantially closer than RS~Oph. For comparable intrinsic high-energy luminosities, its smaller distance would enhance the observed flux by a factor of approximately $3$--$10$ over the commonly adopted RS~Oph distance range. This proximity makes T~CrB an especially favorable target for TeV and multi-messenger observations.

\item The principal uncertainty in predicting the hadronic emission is the density and structure of the circumbinary material. A lower density than in RS~Oph could reduce the efficiency of proton--proton interactions and hence suppress both the hadronic $\gamma$-ray and neutrino signals, even in the presence of a powerful shock. Conversely, a moderate or high effective target density would substantially strengthen the prospects for neutrino detection.

\item We also reassessed the recurrence time of T~CrB. Using the cycle-to-cycle scatter measured among better-sampled recurrent novae, we infer a representative population-dominated recurrence time-scale of $T_{\rm rec}\simeq80\pm12$~yr. If substantial weight is instead assigned to the proposed historical eruptions of T~CrB, and the 1217--1787 interval is assumed to contain seven recurrence cycles, the resulting recurrence history implies a much narrower characteristic time-scale of $\simeq79$--$80$~yr with a dispersion of only $\sim1$--$2$~yr. These values should be regarded as alternative recurrence priors rather than formal predictions of the eruption date, particularly now that the nominal recurrence epoch has been reached without an eruption.

\item Direct cosmic-ray detection from T~CrB is unlikely because charged particles will be deflected by Galactic magnetic fields. A combined TeV-plus-neutrino detection, however, would provide the strongest evidence yet that nova shocks efficiently accelerate hadrons.

\end{itemize}

Our overall conclusion is therefore cautious but optimistic: T~CrB is likely to be at least as promising a target as RS~Oph for TeV detection and may be one of the best nova neutrino targets to date. If its next eruption is detected in both TeV photons and neutrinos, it will become the first genuine multi-messenger nova.

\section*{Acknowledgements}

EA \& KVS Acknowledges NASA award 80NSSC26K0867. LC acknowledges NSF awards AST-2107070 and AST-2205631 and NASA awrads 80NSSC23K0497, 80NSSC25K7334, and 80NSSC26K0601.

BDM acknowledges support from NASA (grants 80NSSC22K0807, 80NSSC24K0408, and 80NSSC26K0300).  The Flatiron Institute is supported by the Simons Foundation. 


\section*{Data availability}
No data is included in this paper.

\bibliographystyle{mnras_vanHack}
\bibliography{biblio}

\appendix

\renewcommand\thetable{\thesection.\arabic{table}}    
\renewcommand\thefigure{\thesection.\arabic{figure}}   
\setcounter{figure}{0}

\section{Supplementary tables}
\label{appB}
In this Appendix we present supplementary tables.

\begin{table*}
\centering
\caption{Observed eruption histories and recurrence-time statistics for the
recurrent-nova sample. Only directly observed eruptions are included in the
primary eruption sequences; inferred or proposed missed eruptions are excluded.
For U Sco and RS Oph, the recurrence statistics used in the population analysis
are calculated from the better sampled modern subsets indicated below.
The dispersion $s_T$ is the sample standard deviation of the inter-eruption
intervals, while $\sigma_{\rm MAD}=1.4826\,{\rm MAD}$ is a robust estimator of
the dispersion. Systems with only one measured interval have no empirical
estimate of the cycle-to-cycle dispersion.}
\label{tab:rn_recurrence}
\small
\begin{tabular}{l p{6.1cm} c c c c c}
\hline
Nova &
Observed eruption epochs &
$N_{\rm int}$ &
$\langle T_{\rm rec}\rangle$ &
$T_{\rm med}$ &
$s_T$ &
$\sigma_{\rm MAD}$ \\
&
&
&
(yr) &
(yr) &
(yr) &
(yr) \\
\hline

M31N 2008-12a &
2008.98, 2009.92, 2010.88, 2011.81, 2012.80,
2013.91, 2014.75, 2015.66, 2016.95, 2017.999,
2018.85, 2019.85, 2020.83, 2021.87, 2022.92,
2023.93 &
15 & 0.996 & 0.992 & 0.110 & 0.086 \\

M31N 2017-01e &
2012.03, 2014.46, 2017.08, 2019.73, 2022.23, 2024.60 &
5 & 2.515 & 2.450 & 0.107 & 0.039 \\

U Sco$^{a}$ &
1863, 1906, 1917, 1936, 1945, 1969, 1979,
1987, 1999, 2010.08, 2022.43 &
5 & 10.69 & 11.08 & 1.76 & 1.60 \\

RS Oph$^{b}$ &
1898, 1907, 1933, 1945, 1958, 1967, 1985,
2006, 2021.60 &
4 & 15.90 & 16.80 & 5.10 & 4.00 \\

T Pyx &
1890, 1902, 1920, 1944, 1967, 2011 &
5 & 24.20 & 23.00 & 12.05 & 7.41 \\

CI Aql &
1917, 1941, 2000 &
2 & 41.50 & 41.50 & 24.75 & \nodata \\

V745 Sco &
1937, 1989, 2014 &
2 & 38.50 & 38.50 & 19.09 & \nodata \\

V3890 Sgr &
1962, 1990, 2019 &
2 & 28.50 & 28.50 & 0.71 & \nodata \\

V394 CrA &
1949, 1987 &
1 & 38.0 & 38.0 & \nodata & \nodata \\

IM Nor &
1920, 2002 &
1 & 82.0 & 82.0 & \nodata & \nodata \\

V2487 Oph &
1900, 1998 &
1 & 98.0 & 98.0 & \nodata & \nodata \\

T CrB &
1866, 1946 &
1 & 79.75 & 79.75 & \nodata & \nodata \\

\hline
\end{tabular}

\begin{flushleft}
\footnotesize
$^{a}$Statistics for U Sco are calculated from the directly observed
1969--2022 sequence. The proposed 2016 eruption is not included.\\
$^{b}$Statistics for RS Oph are calculated from the 1958--2021 sequence
to reduce the effect of heterogeneous early historical coverage.\\
The MAD estimator is not quoted for systems with only two recurrence
intervals because it provides little useful information on the underlying
distribution. Historical candidate eruptions of T CrB are considered
separately in the text and are not included in the primary table.
\end{flushleft}
\end{table*}
\end{document}